\documentclass[aps,prd,twocolumn,fleqn,superscriptaddress]{revtex4}
\usepackage{graphicx,color,natbib}
\usepackage{amsmath,amssymb,amsfonts}

\newcommand{\bse}{\begin{subequations}}
\newcommand{\ese}{\end{subequations}}
\newcommand{\be}{\begin{equation}}
\newcommand{\ee}{\end{equation}}
\newcommand{\bea}{\begin{eqnarray}}
\newcommand{\eea}{\end{eqnarray}}
\newcommand{\ba}{\begin{array}}
\newcommand{\ea}{\end{array}}

\input amssym.def
\input amssym.tex

\usepackage[colorlinks=true, linkcolor=blue, bookmarks=true]{hyperref}

\begin{document}
IPM/P-2015/050

\title{Evolution of Wilson Loop in Time-Dependent ${\cal{N}}=4$ Super Yang-Mills Plasma}

\author{M. Ali-Akbari\footnote{$\rm{m}_{-}$aliakbari@sbu.ac.ir}}
\affiliation{Department of Physics, Shahid Beheshti University G.C., Evin, Tehran 19839, Iran}
\author{F. Charmchi\footnote{charmchi@ipm.ir}}
\affiliation{School of Particles and Accelerators, Institute for Research in Fundamental Sciences (IPM),
P.O.Box 19395-5531, Tehran, Iran}
\author{A. Davody\footnote{davody@ipm.ir}}
\affiliation{School of Particles and Accelerators, Institute for Research in Fundamental Sciences (IPM),
P.O.Box 19395-5531, Tehran, Iran}
\author{H. Ebrahim\footnote{hebrahim@ut.ac.ir}}
\affiliation{Department of Physics, University of Tehran, North Karegar Ave., Tehran 14395-547, Iran}
\affiliation{School of Physics, Institute for Research in Fundamental Sciences (IPM),
P.O.Box 19395-5531, Tehran, Iran}
\author{ L. Shahkarami\footnote{l.shahkarami@du.ac.ir}}
\affiliation{School of Physics, Damghan University, Damghan, 41167-36716, Iran}

\begin{abstract}
Using holography we have studied the evolution of Wilson loop of a quark-antiquark pair in a dynamical strongly coupled plasma. The time-dependent plasma, whose dynamics is originated from the energy injection, is dual to AdS-Vaidya background. The quark-antiquark pair is represented by the endpoints of a string stretched from the boundary to the bulk.  The evolution of the system is studied by evaluating the expectation value of the Wilson loop, throughout the process. Our results show that the evolution of Wilson loop depends on the speed of injecting energy as well as the final temperature of the plasma. For high enough temperatures and rapid energy injection, it starts oscillating around its equilibrium value, immediately after the injection.

\end{abstract}

\maketitle

\tableofcontents

\section{Introduction}
Quark-Gluon Plasma (QGP) is produced at RHIC and LHC by colliding heavy ions, such as gold and lead, at relativistic speeds. Experimental results approve the idea that the plasma is strongly coupled \cite{Shuryak:2003xe,Shuryak:2004cy, CasalderreySolana:2011us}. To describe the properties of such strongly coupled plasma the usual techniques such as lattice gauge theory seem to be not adequate, especially when it comes to real time calculations. An appropriate candidate to study strongly coupled gauge theories with holographic dual is gauge/gravity duality \cite{Natsuume:2014sfa, CasalderreySolana:2011us, Maldacena}.  A well-known example of this duality is AdS/CFT correspondence in which string theory on $AdS_5 \times S^5$ is dual to ${\cal{N}}=4$ SYM theory in four dimensions. In the holographic picture the vacuum (thermal state) in gauge theory is dual to pure AdS (AdS-black hole) background. 

In addition to quarks and gluons, there are stable mesons living in QGP. The potential between quark and antiquark ($Q \bar{Q}$) in QGP at zero as well as non-zero temperature has been evaluated, in the static background \cite{Maldacena:1998im, Brandhuber:1998bs}. The result has been obtained using the holographic picture where the heavy mesons are explained by a string stretched inside the bulk with both endpoints on the boundary. The endpoints of the string represent quark and antiquark. 

At the early stages of heavy ion collision the system is out-of-equilibrium. Studying such time-dependent systems have attracted a lot of attention in the last decade \cite{Chesler:2008hg, Heller:2013oxa}. Time-dependence and strong coupling nature of the system are two obstacles that make describing it even more difficult. Interestingly in the holographic framework people have been able to tackle the problem, to some extent, and explain the process of plasma formation in the gauge theory by black hole formation in the gravity dual \cite{Chesler:2008hg}. In this paper we identify this process by Vaidya metric in the bulk dual and study the evolution of Wilson loop in such time-dependent background. 

\section{Review on String in Static Background}\label{Review}

It is well-known that the static potential energy between a quark and an antiquark is found by evaluating 
the Wilson loop on a closed path, ${\cal{C}}$, where the length of time direction is much larger than the distance between the quarks, ${\cal{T}}\gg l$ \footnote{The Wilson loop has been also applied in other contexts such as \cite{Ghodrati:2015rta}.} \cite{CasalderreySolana:2011us, Maldacena:1998im}. In this limit the Wilson loop takes the following form 
\be%
\label{mass}
\langle W({\cal{C}}) \rangle  =e^{-i(2 m + V(l)){\cal{T}}},
\ee%
where $m$ is the rest  mass of the quarks and $V(l)$ is the potential energy between them. In the strong-coupling regime, holographic picture can be applied to evaluate potential energy of quark-antiquark pair. According to holographic dictionary the expectation value of Wilson loop is dual to the on-shell action of a string with its endpoints separated by distance $l$. In other words \cite{Maldacena:1998im}
\be%
\langle W({\cal{C}})\rangle =e^{i S({\cal{C}})}.
\ee%
$S$ is the Nambu-Goto action describing dynamics of the string in an arbitrary background which is given by 
\be %
\label{action}
 S=\frac{-1}{2 \pi \alpha'} \int_{{\cal{C}}} d\tau d\sigma  \sqrt{- \det(g_{ab})}.
\ee %
where the induced metric is defined by $g_{ab}=G_{MN}\partial_a X^M \partial_b X^N$. $X^M$ ($x^a$) denotes the space-time (world-sheet) coordinates and $G_{MN}$ is the background metric. 

\begin{figure}
\begin{center}
\includegraphics[width=80mm]{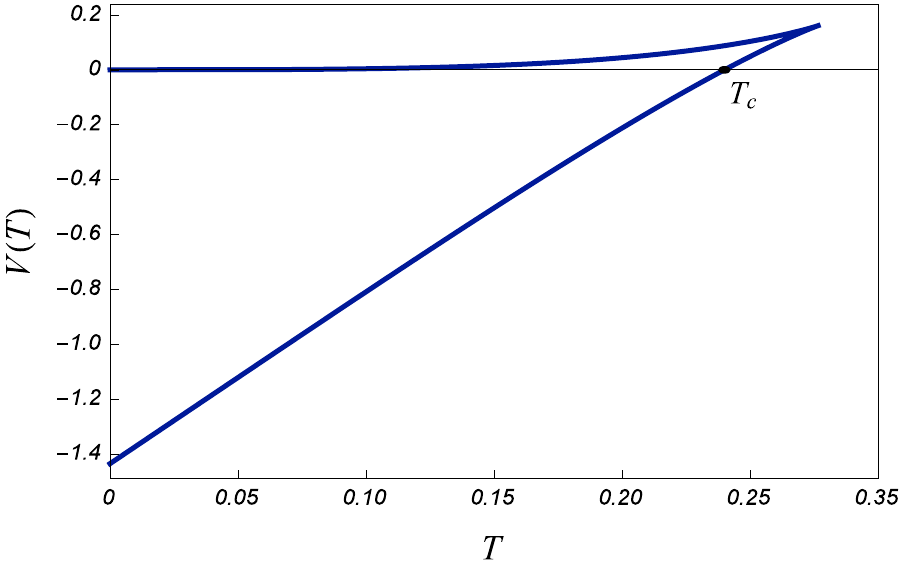}
\caption{The static potential energy between the quark-antiquark pair with respect to temperature. The negative value of the potential energy indicates that the pair is stable. $T_c$ is the temperature of the plasma at which the quark-antiquark pair becomes unstable and dissociates. $l$ is assumed to be one.}\label{VT}
\end{center}
\end{figure}

Using the above idea the potential energy of quark-antiquark pair living in the ${\cal{N}}=4$ quark-gluon plasma has been extensively studied in the literature. The holographic dual of the plasma is described by the AdS-Shwartschild black hole background and a heavy meson is identified with a string hanging from the boundary into the bulk with its endpoints on the boundary. 
The AdS-Shwartschild black hole metric is given by
\be %
 ds^2=\frac{1}{z^2}\left(-f(z) dt^2 + f(z)^{-1} dz^2 + d{\bf{x}}_3^2 \right) + d\Omega_5^2,~~~~~
\ee %
where $f(z)=1-(\frac{z}{z_h})^4$. $z$ is the radial direction and the boundary is located at $z=0$. The horizon of black hole is represented by $z_h$ and its Hawking temperature is $T=\frac{1}{\pi z_h}$ which is identified with the  temperature of plasma. Field theory is living on $(t,\bf{x}_3)$ coordinates. 
If we parametrize the world-sheet coordinate as $\tau=t, \sigma=x$, the shape of string is described by $z(x)$
in the static background. Solving the equation of motion with the boundary conditions $z(\pm \frac{l}{2})=0$, 
we can find the profile of the string. Knowing  the string profile, it is straightforward to compute the on-shell action. Therefore the potential energy between the quark pair can be computed as a function of distance between them. See \cite{CasalderreySolana:2011us, Maldacena:1998im, Brandhuber:1998bs}  for more details.
Here we have plotted the potential energy in terms of the temperature of quark-gluon plasma in figure \ref{VT} . As it is clearly seen in this figure, there exists a critical temperature, $T_c$, at which the potential energy vanishes. 
The meson will be dissociated when the temperature of the plasma is larger than $T_c$, or in  other words,   
the thermal energy overcomes the binding energy between the quarks. This happens because there is another configuration made of two disconnected straight strings stretched between the boundary and the horizon with less energy.  
  
The point we would like to emphasize is  how to compute and subtract the rest mass contribution 
from the on-shell action. The rest mass of the quark is equivalent to  the energy of a straight string stretched between the boundary and the
horizon which is given by 
\be%
\label{thermalmass}
m=\frac{\sqrt{\lambda}}{2 \pi} \int_{\epsilon}^{z_h} \frac{dz}{z^2},
\ee%
where $\lambda$ and $\epsilon$ are t'Hooft coupling constant and UV cut-off, respectively \cite{CasalderreySolana:2011us}. As $\epsilon$ goes to zero, $m$ becomes infinite. The same infinity also appears in the on-shell action due to the UV cut-off. Subtracting the rest mass term from the on-shell action gives the finite physical potential energy, $V(l)$. 

In the time-dependent set-up we need to clarify more on the method of calculating the evolution of Wilson loop:
\begin{itemize}
\item  The condition necessary to define the potential in the static case (${\cal{T}}\gg l$) is not valid in the time-dependent scenario and the relation \eqref{mass} is not trustable. In this case we use
\be%
\label{EWL}
\langle W({\cal{C}})\rangle =e^{-i \int  dt {\cal{W}}(l,t)},
\ee%
which reduces to \eqref{mass} in static situation. ${\cal{W}}(l,t)$ is the on-shell action with the difference that the time coordinate is not integrated over. Similar to the static case due to the UV cut-off ($z\rightarrow 0$) an infinity appears in ${\cal{W}}(l,t)$. In order to regularize ${\cal{W}}(l,t)$ we subtract $m$ \eqref{thermalmass} as follows
\bea%
{\cal{W}}_R(l,t) &=&  {\cal{W}}(l,t) - 2 m \\
&=& \int  d\sigma \left(\sqrt{- \det (g_{ab})}\right)_{on-shell} - 2 m. \nonumber
\eea%
Therefore ${\cal{W}}(l,t)$, with the divergent part removed, is ${\cal{W}}_R(l,t)$.

\item By the evolution of Wilson loop we mean studying the evolution of the reqularized function ${\cal{W}}_R(l,t)$. We will show in the following sections that ${\cal{W}}_R(l,t)$  oscillates about  the value of the static potential with the same parameters after the energy injection is over.
\end{itemize}
 
\section{String in the Vaidya Background}
In this section our goal is to find out how ${\cal{W}}_R(l,t)$ changes when the temperature of the plasma increases. On the gravity side, this is equivalent to studying a hanging string in the Vaidya background. Vaidya metric explains the black hole formation in the bulk due to the collapse of matter, indicating an increase in the plasma temperature due to the energy injection, via an out-of-equilibrium process. While the temperature changes, at each time, we calculate ${\cal{W}}_R(l,t)$.

The Vaidya metric in Eddington-Finkelstein coordinates is
\be%
ds^2 = \frac{1}{z^2} (-F(V,z) dV^2 - 2 dV dz + d{\bf{x}}_3^2) + d\Omega_5^2,~~~~
\ee%
where 
\be%
F(V,z) = 1-M(V) z^4,
\ee%
and $V$ is the time direction on the boundary. $M(V)$ is the mass function of the black hole. It starts from zero and reaches its finite value in the time interval of energy injection and remains constant afterwards.

In order to find the time-dependent ${\cal{W}}_R(l,t)$, similar to the calculation done in \cite{Ishii:2014paa, Ishii:2015wua, Ali-Akbari:2015bha}, it is helpful to introduce the null coordinates  $(u,v)$ on the string worldsheet. Thus the background coordinates will depend on $(u,v)$ and apart from  $V(u,v)$, $Z(u,v)$ and $X(u,v)$, we set all other coordinates to zero. Using the Numbo-Goto action \eqref{action} the equations of motion for the mentioned fields become

\bse\label{eom}\begin{align}%
V_{,uv}&=
\left( \frac{F_{,Z} }{2} -\frac{F}{Z}\right) V_{,u} V_{,v} + \frac{1}{Z} X_{,u} X_{,v}\, ,
\\
Z_{,uv}&=
\left(\frac{F^2}{Z} - \frac{F }{2} F_{,Z} -\frac{1}{2} F_{,V}\right) V_{,u} V_{,v}  \nonumber\\
&+ \left(\frac{F}{Z} -\frac{F_{,Z}}{2}\right) \left(Z_{,u} V_{,v}+Z_{,v} V_{,u}\right)\cr
&+ \frac{2}{Z} Z_{,u} Z_{,v}   - \frac{F}{Z} X_{,u} X_{,v}\, ,\\
X_{,uv}& = \frac{Z_{,u} X_{,v} + Z_{,v} X_{,u}}{Z}.
\
\end{align}\ese%
Notice that since we would like the $u$ and $v$ coordinates to stay null, we need to impose the following constraint equations, coming from $g_{uu}=0$ and $g_{vv}=0$,
\bse\begin{align}%
\label{cons1} C_1 = \frac{1}{Z^2} (F(V,Z) V_{,u}^2 + 2 V_{,u}  Z_{,u} - X_{,u}^2) & = 0,\\
\label{cons2} C_2 = \frac{1}{Z^2} (F(V,Z) V_{,v}^2 + 2 V_{,v}  Z_{,v} - X_{,v}^2 ) & = 0.
\end{align}\ese%
To determine the dynamics of the string in the Vaidya background, we will numerically solve the equations of motion. One can see that $\partial_v C_1 = \partial_u C_2 = 0$ on-shell. Provided the initial and boundary conditions are consistent with the constraint equations, this will guarantee that the constraint equations will be satisfied during the evolution. Thus we need to know the appropriate boundary and initial conditions as we will summarize them in the following subsections. 

\subsection{Boundary Conditions}
In order to solve the equations of motion and obtain the physical observables, we have to impose the appropriate boundary conditions on the AdS boundary. Notice that since the two endpoints of the string are on the boundary we have to impose two sets of boundary conditions, one for each endpoint. According to the discussions in \cite{Ishii:2014paa}, by fixing the diffeomorphism on the string worldsheet one may choose the AdS boundary to be at $u=v$ for one of the points and $u=v+L$ for the other one. The boundary condition on $Z$ and $X$ are simply
\bea%
Z|_{u=v}&=&0~;~~~~X|_{u=v}=\frac{-l}{2},\\
Z|_{u=v+L}&=&0~;~~~~X|_{u=v+L}=\frac{l}{2}.
\eea
In order to obtain the rest of the boundary conditions we expand the fields near the boundary ($u=v$) as 
\bea%
V(u,v)&=&V_0(v) + V_1(v) (u-v) + ...~,~~~~~~~~\\  
Z(u,v)&=&Z_1(v) (u-v) +Z_2(v) (u-v)^2+ ...~,~~~~~~~~\\
X(u,v)&=&\frac{-l}{2} + X_1(v) (u-v) + ...~,~~~~~~~~
\eea%
and substitute them into the evolution equations. We do the same for the other boundary ($u=v+L$) and get similar results. By imposing the regularity condition at $u=v$ and $u=v+L$ we can figure out the remaining boundary conditions. We also have to check the consistency of the results with the constraint equations. To summarize we get the following boundary conditions at $u=v$:
\bse\begin{align}%
\label{bc1}
V(u,v)&=V_0(v) + {\cal{O}}\left((u-v)^5\right),\\ \nonumber
\label{bc2}  
Z(u,v)&=\frac{\dot{V}_0(v)}{2} (u-v) + \frac{\ddot{V}_0(v)}{4} (u-v)^2\\ 
&+\frac{\dddot{V}_0(v)}{12} (u-v)^3 + {\cal{O}}\left((u-v)^4\right),\\ 
X(u,v)&=\frac{-l}{2} + {\cal{O}}\left((u-v)^3\right), 
\end{align}\ese%
which imply
\be%
Z_{,uv}|_{u=v} = 0,\ 2 Z_{,u}|_{u=v} = \dot{V}_0(v).
\ee%
Similar calculations can be done for $u=v+L$. We refer the interested reader to \cite{Ishii:2014paa} for more details.

\subsection{Initial Conditions}
The space-time background is dynamical and evolves from pure AdS to AdS-black brane. Thus, for $V < 0$,  the static solution of the string hanging in the bulk of pure AdS can be set as the initial condition. This solution has been previously obtained in the literature, for instance see \cite{CasalderreySolana:2011us} and references therein. We will redo this calculation in EF coordinate system. We choose the string worldsheet coordinates $(\tau,\sigma)$ to lie along $(V,X)$ and $Z$, which describes the shape of the string, to depend on $X$. Therefore the action of the string becomes
\be%
S = \frac{-1}{2\pi\alpha'} \int_{\frac{-l}{2}}^{\frac{l}{2}} dV dX \frac{1}{Z^2} \sqrt{1+{Z'}^2},
\ee%
where $Z'=\frac{dZ}{dX}$. Since the Lagrangian does not explicitly depend on $X$ we can use the associated Hamiltonian, which is the constant of motion, to obtain the static solution. So we get
\be%
\label{static solution}
\frac{dZ}{dX} = -\frac{Z_*^2}{Z^2} \sqrt{1-\frac{Z^4}{Z_*^4}}.
\ee%
$Z_*$ is where $\frac{dZ}{dX}=0$. The minus sign has been chosen here since we are solving the equations for the half of the string where $\frac{-l}{2}<X<0$. 

To obtain the initial conditions for the variables $V$, $Z$ and $X$ we use the constraint equations and the static solution above, \eqref{static solution}. Note that since $V_{,v}>0$ at the boundary, therefore by using the boundary conditions \eqref{bc1} and \eqref{bc2}, $Z_{,u}>0$ and $Z_{,v}<0$. Applying these conditions on $Z$ and $V$ derivatives and  using $X_{,u}|_{Z=0}=X_{,v}|_{Z=0}=0$, the constraint equations \eqref{cons1} and \eqref{cons2} lead to
\bea%
\label{eqV1}
V_{,u} &=& - Z_{,u} + \sqrt{Z_{,u}^2 + X_{,u}^2} \cr
&=& Z_{,u} \bigg(-1+\sqrt{1+\big(\frac{dX}{dZ}\big)^2}\bigg) ,~\\
\label{eqV2}
V_{,v} &=& - Z_{,v} + \sqrt{Z_{,v}^2 + X_{,v}^2} \cr
&=& Z_{,v} \bigg(-1-\sqrt{1+\big(\frac{dX}{dZ}\big)^2}\bigg).
\eea%
If we calculate the second derivative of the above equations with respect to $v$ and $u$, respectively and set them equal to each other we obtain
\be%
\label{master}
Z_{,uv} \bigg( \sqrt{1+\big(\frac{dX}{dZ}\big)^2}\bigg) + Z_{,v} Z_{,u} \bigg(\sqrt{1+\big(\frac{dX}{dZ}\big)^2}\bigg)_{,Z}=0.~~~~~
\ee%
We can recast this equation as 
\be%
\bigg( Z_{,u} \sqrt{1+\big(\frac{dX}{dZ}\big)^2}  \bigg)_{,v}=0.
\ee%
The above equation can be solved using \eqref{static solution}. The final result is
\be%
\label{solZ}
Z \, _2F_1\bigg(\frac{1}{4},\frac{1}{2};\frac{5}{4};\frac{Z^4}{\text{Z}_*^4}\bigg) = \phi(u) - \phi(v),
\ee%
where $\phi$ is an arbitrary function. The sign on the right hand side is fixed to have the left hand side equal to zero at $u=v$ or $Z=0$. 

By integrating \eqref{static solution} the initial condition on $X(u,v)$ becomes
\be%
X(u,v) = \frac{l}{2} - \frac{Z^3}{3 \text{Z}_*^2} \, _2F_1\left(\frac{1}{2},\frac{3}{4};\frac{7}{4};\frac{Z^4}{\text{Z}_*^4}\right),
\ee%
and using the fact that at $Z=Z_*$, $X=0$ we get
\be%
Z_*=\frac{3 \Gamma \left(\frac{5}{4}\right)}{2 \sqrt{\pi } \Gamma \left(\frac{7}{4}\right)} l.
\ee%
The initial condition on $V$ is obtained from \eqref{eqV1} and \eqref{eqV2} 
\bea%
V(u,v)&=&- Z \left(1-\, _2F_1\left(\frac{1}{4},\frac{1}{2};\frac{5}{4};\frac{Z^4}{\text{Z}_*^4}\right)\right)+\chi(v),~~~~~~~\\
V(u,v)&=&- Z \left(1+\, _2F_1\left(\frac{1}{4},\frac{1}{2};\frac{5}{4};\frac{Z^4}{\text{Z}_*^4}\right)\right)+{\tilde{\chi}}(u),~~~~~~~
\eea%
where $\chi$ and $\tilde{\chi}$ are arbitrary functions. If we set the above equations equal to each other and use \eqref{solZ} we obtain these functions as
\be%
\chi(v) = 2 \phi(v)~,~~~~~{\tilde{\chi}}(u) = 2 \phi(u).
\ee%
How to fix the arbitrary function $\phi$ has been extensively discussed in \cite{Ishii:2014paa} which we will not go through the details here. In our case we can choose $\phi(y) = y$. 
\section{Numerical Results}
So far we have prepared the necessary ingredients to solve the equations of motion in the Vaidya background. The time-dependent function $M(V)$ has been selected as
\bea %
\label{M}
 M(V)= M_f \left\{%
\begin{array}{ll}
    0 & V<0, \\
    \frac{1}{2}\left[1-\cos(\frac{\pi V}{\Delta V})\right] & 0 \leq V \leq \Delta V, ~~~~ \\
   1 & V>\Delta V ,\\
\end{array}%
\right. \eea %
where $\Delta V$ is the time interval in which the mass of the black hole increases from zero to the final value $M_f$ which is constant. Note that the radius of the event horizon is $z_h=M_f^{\frac{-1}{4}}$.

The equations of motion \eqref{eom} are solved in the above time-dependent background. At each instance of time we calculate the ${\cal{W}}_R(l,t)$.
\begin{figure}
\begin{center}
\includegraphics[width=80mm]{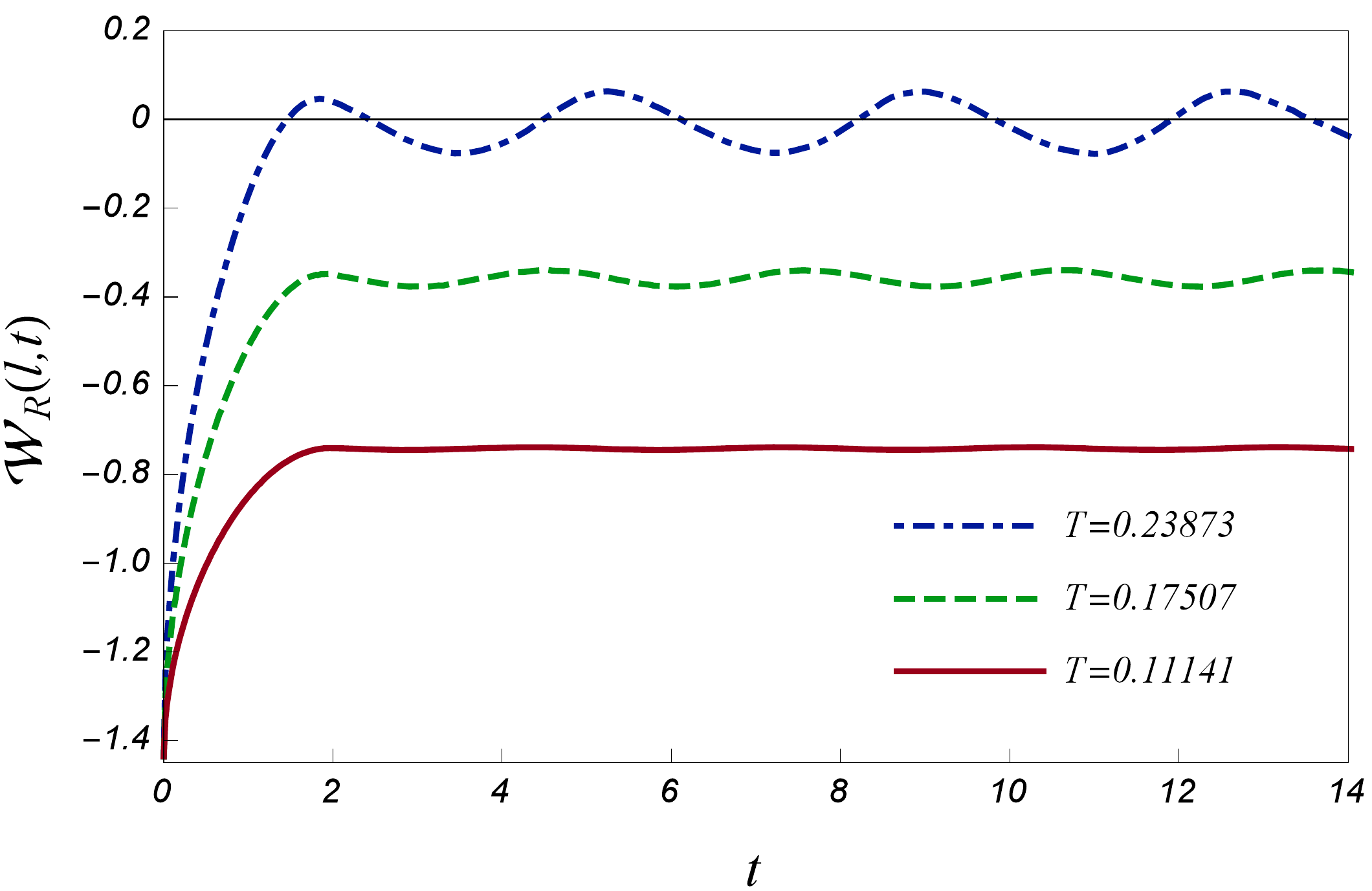}
\caption{${\cal{W}}_R(l,t)$ with respect to the boundary time for different values of the temperature. In this figure we have set $\Delta V = 2$ and $l=1$. The corresponding static potentials from top to bottom are $V(l) =-0.0065104, -0.355009, -0.738675$. ${\cal{W}}_R(l,t)$ in each curve oscillates around the corresponding static potential.}\label{Vtdeltav2}
\end{center}
\end{figure}

Assuming fixed values of $l$ and $\Delta V$, we have plotted the time-dependent ${\cal{W}}_R(l,t)$ for different values of the final temperature in figure \ref{Vtdeltav2}. At low temperature (red curve) the turning point of the string, $Z_*$, is far away from the final horizon and therefore ${\cal{W}}_R(l,t)$ reaches approximately a constant value and stays there afterwards. By increasing the final temperature of the plasma (green and blue curves) the ${\cal{W}}_R(l,t)$ starts oscillating around its equilibrium value more noticeably. Since the temperature is higher than the previous case, the turning point of the string is closer to the horizon and therefore the energy injection causes the ${\cal{W}}_R(l,t)$ to oscillate with larger amplitude.
\begin{figure}
\begin{center}
\includegraphics[width=80mm]{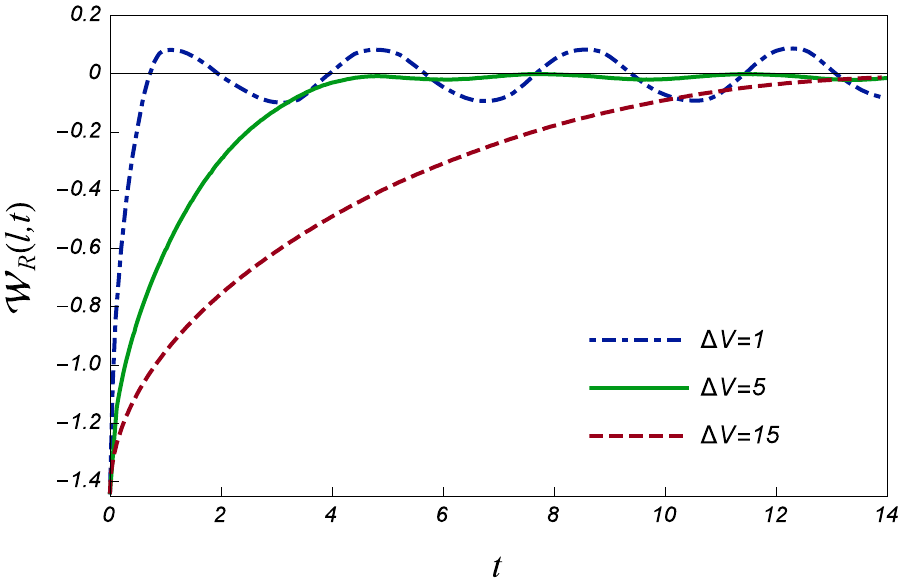}
\caption{${\cal{W}}_R(l,t)$ with respect to the boundary time for different values of energy injection time interval. In this figure we have set $T = 0.23873$ and $l=1$. Raising $\Delta V$ decreases the energy injection rate. The corresponding static potential is $V(l) =-0.0065104$.}\label{VtTfixed}
\end{center}
\end{figure}

The reason for these oscillations can be explained as follows. At $t=0$ the meson bound-state which is stable and in its ground state is described by the static string stretched in the bulk with the end points on the boundary. As the temperature is raised the shape of the brane changes time-dependently. The turning point of the string goes closer to the horizon as the energy is being injected. Our numerical calculations show that the string oscillates about the string static solution corresponding to the final temperature of the quench. In the field theory description these oscillations present themselves as the oscillations in the time-dependent Wilson loop, ${\cal{W}}_R(l,t)$. These oscillations can be interpreted as if the quench puts the meson in the excited state. This conclusion is reasonable regarding to the result of \cite{Ishii:2014paa} where the authors have studied the dynamics of the shape of the brane in a time-dependent background. The power spectrum of the condensation oscillations gives the excited mesonic modes in the field theory dual. Note that in the large $N$ limit that we have done the calculations in this paper, the excited meson is stable since there is no energy dissipation. 

In figure \ref{VtTfixed}, ${\cal{W}}_R(l,t)$ at fixed values of $l$ and $T$ has been plotted for various speeds of energy injection. The final temperature has such value that the turning point of the string is close to the horizon. Therefore by rapid energy injection ($\Delta V=1$) ${\cal{W}}_R(l,t)$ starts oscillating with large amplitude immediately after the injection is over. Decreasing the speed of the energy injection delays its oscillations with smaller amplitude. We should emphasize that the oscillations in all cases happen around the negative equilibrium value of static potential.  

Our calculations show that two independent parameters affect the response of the quark pair to the energy injection. One, as we expected, is the final temperature of the plasma and the other is the injection speed. Regarding figure \ref{Vtdeltav2} we explained that the meson excitations depend on the final temperature of the plasma when the rate of the injection is held fixed. However figure \ref{VtTfixed} shows that for the same final temperature, the excitations are dependent on the rate of the energy injection. Therefore excitations of the meson in the plasma seem to depend on the temperature as well as the energy injection rate.      

{\bf{{Acknowledgment}}}
We would like to thank the referee for illuminating comments which improved the quality of the paper. H. E. would like to acknowledge the financial support of University of Tehran for this research under the grant number 392692/1/01.

\end{document}